\documentclass[apj]{emulateapj}

\shortauthors{{\sc Rasmussen et al.}}

\begin{document}

\title{The Suppression of Star Formation and the Effect of Galaxy
  Environment in Low-Redshift Galaxy Groups$^\star$}

\author{Jesper Rasmussen,\altaffilmark{1} John S.~Mulchaey,\altaffilmark{2} 
  Lei Bai,\altaffilmark{3} Trevor J.~Ponman,\altaffilmark{4} 
  Somak Raychaudhury,\altaffilmark{4} and Ali Dariush\altaffilmark{5}}

\altaffiltext{1}{Dark Cosmology Centre, Niels Bohr Institute,
  University of Copenhagen, Juliane Maries Vej 30, DK-2100 Copenhagen,
  Denmark; jr@dark-cosmology.dk}

\altaffiltext{2}{Carnegie Observatories, 813 Santa Barbara Street,
  Pasadena, CA 91101, USA}

\altaffiltext{3}{Department of Astronomy and Astrophysics, University
  of Toronto, 50 St.\ George Street, Toronto, Ontario, M5S 3H4,
  Canada}

\altaffiltext{4}{School of Physics and Astronomy, University of
  Birmingham, Edgbaston, Birmingham B15 2TT, UK}

\altaffiltext{5}{Physics Department, Imperial College London, Prince
  Consort Road, London SW7 2AZ, UK}

\altaffiltext{$\star$}{This paper includes data gathered with the 6.5 meter
Magellan Telescopes located at Las Campanas Observatory, Chile.}

\begin{abstract}
  Understanding the interaction between galaxies and their
  surroundings is central to building a coherent picture of galaxy
  evolution. Here we use {\em GALEX} imaging of a statistically
  representative sample of 23 galaxy groups at $z\approx 0.06$ to
  explore how local and global group environment affect the UV
  properties and dust-corrected star formation rates of their member
  galaxies. The data provide star formation rates out to beyond
  $2R_{200}$ in all groups, down to a completeness limit and limiting
  galaxy stellar mass of $0.06$~$M_\odot$~yr$^{-1}$ and $1\times
  10^8$~$M_\odot$, respectively. At fixed galaxy stellar mass, we find
  that the fraction of star-forming group members is suppressed
  relative to the field out to an average radius of $R\approx
  1.5$~Mpc\,$\approx 2R_{200}$, mirroring results for massive
  clusters. For the first time we also report a similar suppression of
  the specific star formation rate within such galaxies, on average by
  40\% relative to the field, thus directly revealing the impact of
  the group environment in quenching star formation within infalling
  galaxies. At fixed galaxy density and stellar mass, this suppression
  is stronger in more massive groups, implying that both local and
  global group environment play a role in quenching. The results favor
  an average quenching timescale of $\ga 2$~Gyr and strongly suggest
  that a combination of tidal interactions and starvation is
  responsible. Despite their past and ongoing quenching, galaxy groups
  with more than four members still account for at least $\sim 25$\%
  of the total UV output in the nearby universe.
\end{abstract}

\keywords{galaxies: evolution --- galaxies: groups: general ---
  galaxies: star formation --- ultraviolet: galaxies}

\section{Introduction}\label{sec,intro}

Galaxies may evolve from star-forming late-types to passive
early-types either via internal, secular processes or through
mechanisms induced by their environment. Disentangling these two
pathways and understanding the nature and impact of various
environmental processes on galaxy evolution is a major goal of
contemporary astrophysics. While the overall star formation density of
the universe has declined significantly since a redshift of $z\approx
2$ \citep{mada96,hopk06}, this evolution appears to be accelerated in
dense environments, with groups and clusters of galaxies containing a
lower fraction of star-forming galaxies at fixed stellar mass and
redshift than the general field (e.g., \citealt{kauf04,mcge11}).

Recent work has demonstrated that the effect of secular evolution in
quenching star formation, primarily driven by galaxy stellar mass, can
be separated from that of the galaxy environment \citep{peng10}. The
former effect dominates at high galaxy masses, whereas environmental
quenching of star formation becomes increasingly important at lower
masses. Possible mechanisms for a more rapid quenching of low-mass
galaxies in dense environments include tidal interactions, harassment,
ram pressure stripping, starvation, and major and minor mergers.

Groups of galaxies are particularly interesting in this respect.
Containing around half of all galaxies and most of the stellar mass in
the local universe \citep{eke04,eke05}, they represent a key
environment in the hierarchical build-up of cosmic structure. The
galaxy population within groups has evolved significantly since
$z\approx 0.5$, with the fraction of emission-line galaxies having
declined substantially \citep{wilm05}. Interestingly, however, the
emerging picture is one in which star formation in star-forming
galaxies of a given stellar mass is similar in groups and the field,
whereas the fraction of such galaxies is not (e.g.,
\citealt{balo04,bald06,iovi10}). For example, \citet{balo04} used
inferred H$\alpha$ equivalent widths of galaxies within a large sample
of low-redshift groups in the Sloan Digital Sky Survey (SDSS) and 2dF
Galaxy Redshift Survey (2dFGRS) catalogs to show that the fraction of
star-forming galaxies depends systematically on local density, but the
actual star formation rate {\em in} such galaxies does not. This has
recently been shown to apply also at intermediate redshifts, $z \sim
0.4$ and at fixed stellar mass \citep{mcge11}, and the result has
further been extended to galaxies in both nearby \citep{hain11} and
distant ($z \sim 1$; \citealt{muzz12}) clusters and superclusters.

This has been interpreted as evidence that the truncation of star
formation is accomplished on short timescales, rapidly enough to leave
the average star formation properties of star-forming galaxies largely
unaffected. This would be true even in groups of fairly low mass, and
support for this comes from recent studies based on UV star formation
rates in Hickson Compact Groups which are generally relatively sparse
systems. Such work has revealed a pronounced deficit of galaxies with
moderate specific (i.e., stellar-mass-normalized) star formation rates
\citep{tzan10,walk12}, consistent with a rapid, environment--driven
transition from (perhaps temporarily enhanced) star formation to
quiescence. Although the dominant responsible mechanism(s) have yet to
be unambiguously identified, it is clear that galaxy--galaxy
interactions could play a prominent role both in compact and more
typical groups, given the relatively low velocity dispersions
characteristic of these environments.

Recognizing that a detailed multi-wavelength characterization of the
coupling between global group properties and those of the group
members themselves could offer important new insights, we initiated
the {\em XI} Groups Survey, a study of a statistically representative
sample of 25 redshift-selected groups at $z\approx 0.06$
\citep{rasm06b}. In this framework, we have previously presented star
formation rates (SFRs) from {\em Spitzer}/MIPS 24\,$\mu$m data for a
subset of nine groups \citep{bai10}. Results revealed star-forming
galaxy fractions that bridge those of the field and massive clusters,
being everywhere higher than those in cluster outskirts but lower than
in the field by $\sim 30$\% on average. The fractions showed no
systematic dependence on global group properties such as velocity
dispersion or total stellar mass. In addition, and in line with the
above previous studies, specific SFRs were generally found to be
similar to those in the field.

One limitation of our {\em Spitzer} study was the narrowness of the
rectangular region ($20\arcmin$, equivalent to $\sim 0.7$~Mpc radius
at the sample redshift) covered by our MIPS observations. To fully
understand the suppression of star formation in groups requires
establishing the star-forming properties of group members out to
larger radii, and including galaxies that are encountering the group
environment for the first time. For example, simulations
\citep{kawa08} and isolated observations \citep{rasm06a} suggest that
gaseous stripping and starvation can suppress star formation in disk
galaxies on their first passage through even low-mass groups. In more
massive clusters, star formation rates appear suppressed out to $R\sim
2R_{200}$ (\citealt{balo98}; here $R_{200}$ is the radius enclosing a
mean overdensity of 200 relative to the critical density). Recent
studies have extended this conclusion to even larger radii
\citep{chun11}, revealing a suppression in the fraction of
star-forming galaxies out to $\sim 7$~Mpc from cluster cores
\citep{lu12}. However, other works have also shown that star formation
can be locally enhanced well outside the cluster virial radius for
galaxies within the filaments feeding these structures
\citep{port07,fadd08,port08,pere10}, with galaxy--galaxy interactions
among infalling galaxies providing a possible explanation. Testing the
general validity of these results for low-mass groups requires full
coverage of the group environment out to the field and infall regions.

Here we extend our previous {\em Spitzer} study by presenting {\em
  GALEX} data for most of the full {\em XI} sample. Apart from
providing improved statistics due to the increased sample size, the
{\em GALEX} field-of-view of $R=36\arcmin$ also allows coverage of
each group field out to $R\sim 2.5$~Mpc, corresponding to at least
$2R_{200}$ for these systems. This continuous coverage of all regions
from the general "field" through the infall regions to the dense group
cores enables a complete census of the star-forming properties of
group galaxies, including those only now encountering the group
environment.

We assume $H_0=72$~km~s$^{-1}$~Mpc$^{-1}$, $\Omega_m=0.27$, and
$\Omega_{\Lambda}=0.73$. At the sample median redshift of $z=0.061$,
$1\arcmin$ corresponds to 69~kpc and the total {\em GALEX} field to
$\approx 5.0$~Mpc diameter. All uncertainties are quoted at the
$1\sigma$ level.

\section{Observations and Analysis}\label{sec,sample}

\subsection{Sample and Optical Observations}

Details of the {\em XI} sample selection are given in \cite{rasm06b},
but we repeat here that the sample comprises 25 groups selected from
systems identified by \citet{merc02} in the 100-K data release of the
2dFGRS \citep{coll01}. Groups with at least five members and having
velocity dispersion $\sigma < 500$~km~s$^{-1}$ were selected randomly
within the redshift range $z=0.060$--0.062 to span the full parameter
space in $\sigma$, galaxy richness, and estimated virial radii.
Barring any inherent biases in the friends-of-friends algorithm used
to build the parent group catalog, the sample is intended to be
unbiased and truly representative of the low-redshift group
population.

An extensive imaging and spectroscopic campaign with the IMACS
spectrograph at the 6.5-m Baade/Magellan telescope at Las Campanas has
more than doubled the number of spectroscopically identified group
members compared to the original \citet{merc02} catalog. This has
enabled highly reliable estimates of velocity dispersion and
luminosity centroid for each group. In the present work, we adopt as
group centers the location of the brightest group galaxy within a
projected distance of 0.5~Mpc of the $R$-band luminosity-weighted
center, and with velocity within $2\sigma$ of the group mean. A
forthcoming paper (L.~Bai et~al., in preparation) will provide the
full details and results of our optical spectroscopy and {\em Spitzer}
24\,$\mu$m imaging of the full {\em XI} sample, along with the
individual galaxy redshifts, stellar mass estimates, UV photometric
measurements, and SFR estimates used in the present work.

Following the procedure in \citet{bai10}, we have here complemented
our optical spectroscopy within the $R\approx 15\arcmin$ IMACS
field-of-view with redshift measurements from existing catalogs out to
$R=35 \arcmin$ from the center of each group. These redshifts come
mainly from the 2dFGRS (see \citealt{bai10} for further details).
Galaxies were included in the present study if having recession
velocities within $3\sigma$ of the group mean, as determined from the
galaxies within the central $R=1$~Mpc. Some fraction of these objects
are likely to represent a population of infalling galaxies that are
only now encountering the group environment, and some may represent
true ``field'' galaxies. This allows us to probe the impact of the
group environment on the UV properties of galaxies out to well beyond
the group virial radii. However, we emphasize that this auxiliary
sample is more heterogeneous than the one covered by our IMACS
spectroscopy, and that some of these galaxies should be considered
candidate rather than bona fide group members. We will comment on the
potential impact of this whenever relevant.

A subset of the group members has five-band photometry from the SDSS.
For these galaxies, we used the {\em kcorrect} package \citep{blan07},
to derive a relation between their $R$--band stellar mass-to-light
ratio and their $B$--$R$ color. This relation was then used to
estimate stellar masses $M_\ast$ for the full group sample, with a
typical uncertainty for individual galaxies of $\sim 45$\%
(cf.\ \citealt{bell03}). For 17 of the 833 galaxies discussed in this
paper (2\%), only $R$ magnitudes are available, and $M_\ast$ for these
was determined in a similar fashion from their $R$ magnitudes alone.

\subsection{{\em GALEX} Observations and Photometry}

Of the 25 groups within the {\em XI} sample, 22 were imaged by {\em
  GALEX} in both bands (NUV: 1770--2830~\AA ; FUV: 1340--1790~\AA ) as
part of GI Cycle 4 (PI: Mulchaey). The median exposure in both bands
was 1600~s. One further group (MZ\,5383) was covered in the {\em
  GALEX} Medium Imaging Survey to comparable depth.  The final two
{\em XI} groups, MZ\,3849 and MZ\,9994, suffer from bright stars
within the field above the allowed {\em GALEX} limit and so are not
considered here. In addition, one of the groups, MZ\,770, was only
partially covered by {\em GALEX}, with the group center lying outside
the {\em GALEX} field-of-view. This group is excluded whenever
considering group-wide quantities derived from the {\em GALEX} data.

The {\em GALEX} exposure times were motivated by the need to match the
depth of our IMACS spectroscopy and {\em Spitzer}/MIPS data, requiring
reliable (signal-to-noise ratio S/N\,$\ga 10$) UV flux measurements
down to $m_{\rm AB}\approx 21.75$ in the NUV band, corresponding to a
star formation rate of $\sim 0.1$~M$_\odot$~yr$^{-1}$. For each group,
the effective NUV and FUV exposure times are identical, except for
MZ\,5383 which has a significantly longer NUV exposure of
2894~s. Table~\ref{tab,log} summarizes the available {\em GALEX}
observations of the 23 group fields.

\begin{deluxetable*}{lcccccccrrrr}
\tabletypesize{\scriptsize}
%\rotate
\tablecaption{Summary of {\em GALEX} Observations and General Properties of the {\em XI} Sample\label{tab,log}}
\tablewidth{0pt}
\tablehead{ \colhead{Group} & \colhead{R.A} & \colhead{Decl.} & 
  \colhead{$t_{\rm  exp}$} & \colhead{Ext.} & \colhead{$\langle z\rangle$} & 
  \colhead{$\sigma$} & \colhead{$M_{\ast,{\rm tot}}$} & \colhead{$N$} & 
  \colhead{$N_{\rm gal}$} & \colhead{$N_{\rm UV}$} & \colhead{$N_{\rm 24\mu m}$} \\
  \colhead{ } & \colhead{(J2000)} & \colhead{(J2000)}  & \colhead{(s)} & 
  \colhead{(mag)} &  \colhead{ } & \colhead{(km~s$^{-1}$)} & \colhead{(log\,$M_\odot$)} & 
  \colhead{ } &  \colhead{ }& \colhead{ }& 
  \colhead{ } \\
\colhead{(1)} & \colhead{(2)} & \colhead{(3)} & \colhead{(4)} & \colhead{(5)} &
\colhead{(6)} & \colhead{(7)} & \colhead{(8)} & \colhead{(9)} & \colhead{(10)}&
\colhead{(11)} & \colhead{(12)}
} 
 \startdata
MZ\,770	      & $22^h 18^m 03\fs 60$  & $-28^\circ 22\arcmin 58\farcs 8$ 
              & 1,245 & 0.14 & 0.0606 & 215 & 11.21 & 15 & 12 &  9 &  2 \\
MZ\,1766      & $00^h 38^m 31\fs 92$  & $-27^\circ 11\arcmin 16\farcs 8$ 
              & 1,658 & 0.11 & 0.0611 & 222 & 11.62 & 15 & 40 & 30 & 10 \\
MZ\,3067      & $22^h 16^m 16\fs 32$  & $-25^\circ 42\arcmin 25\farcs 2$ 
              & 1,333 & 0.18 & 0.0602 & 160 & 11.49 & 16 & 32 & 29 &  8 \\
MZ\,3182      & $22^h 19^m 17\fs 28$  & $-27^\circ 01\arcmin 22\farcs 8$ 
              & 1,563 & 0.15 & 0.0611 & 258 & 11.14 &  8 & 22 & 18 &  5 \\
MZ\,3541      & $10^h 03^m 41\fs 04$  & $-04^\circ 09\arcmin 50\farcs 4$ 
              & 1,586 & 0.30 & 0.0627 & 121 & 11.30 & 11 & 21 & 16 & 10 \\
MZ\,3698      & $09^h 59^m 27\fs 36$  & $-05^\circ 43\arcmin 58\farcs 8$
              & 1,600 & 0.31 & 0.0609 & 215 & 11.53 & 14 & 21 & 12 &  7 \\
MZ\,4001      & $10^h 16^m 23\fs 28$  & $-03^\circ 15\arcmin 18\farcs 0$
              & 1,593 & 0.27 & 0.0593 & 325 & 11.73 & 36 & 60 & 45 & 15 \\
MZ\,4548      & $10^h 53^m 52\fs 80$  & $-05^\circ 59\arcmin 42\farcs 0$ 
              & 1,565 & 0.25 & 0.0625 & 157 & 10.86 & 12 & 24 & 19 &  7 \\
MZ\,4577      & $11^h 33^m 05\fs 04$  & $-04^\circ 00\arcmin 46\farcs 8$ 
              & 2,609 & 0.44 & 0.0621 & 232 & 11.35 & 20 & 44 & 31 &  9 \\
MZ\,4592      & $11^h 30^m 51\fs 12$  & $-03^\circ 47\arcmin 27\farcs 6$ 
              & 1,656 & 0.51 & 0.0616 & 206 & 11.50 & 18 & 37 & 24 & 10 \\
MZ\,4881      & $11^h 39^m 48\fs 00$  & $-03^\circ 30\arcmin 28\farcs 8$
              & 2,885 & 0.19 & 0.0613 & 510 & 12.12 & 34 & 52 & 42 & 21 \\
MZ\,4940      & $11^h 36^m 04\fs 80$  & $-03^\circ 39\arcmin 57\farcs 6$
              & 3,171 & 0.37 & 0.0620 &  64 & 11.12 &  8 & 13 & 11 &  5\\
MZ\,5293      & $12^h 16^m 19\fs 92$  & $-03^\circ 21\arcmin 00\farcs 0$ 
              & 1,666 & 0.33 & 0.0620 & 104 & 11.32 &  9 & 10 &  5 &  4\\
MZ\,5383      & $12^h 35^m 47\fs 52$  & $-03^\circ 41\arcmin 24\farcs 0$ 
              & 1,186 & 0.30 & 0.0595 & 190 & 11.39 & 17 & 31 & 24 &  3\\
MZ\,5388      & $12^h 34^m 04\fs 56$  & $-03^\circ 22\arcmin 22\farcs 8$ 
              & 1,670 & 0.33 & 0.0598 & 144 & 11.33 & 14 & 19 & 16 &  8 \\
MZ\,8816      & $00^h 05^m 58\fs 32$  & $-27^\circ 52\arcmin 26\farcs 4$ 
              & 1,564 & 0.14 & 0.0608 & 291 & 11.48 & 25 & 68 & 34 & 16 \\
MZ\,9014      & $00^h 37^m 48\fs 00$  & $-27^\circ 30\arcmin 28\farcs 8$ 
              & 1,750 & 0.11 & 0.0608 & 283 & 11.49 & 24 & 46 & 40 &  6\\
MZ\,9069      & $00^h 28^m 25\fs 20$  & $-27^\circ 28\arcmin 58\farcs 8$
              & 1,584 & 0.13 & 0.0615 & 361 & 11.43 & 29 & 64 & 45 & 14\\
MZ\,9137      & $00^h 18^m 56\fs 64$  & $-27^\circ 54\arcmin 57\farcs 6$ 
              & 1,596 & 0.19 & 0.0601 & 329 & 11.15 & 13 & 22 & 17 &  6 \\
MZ\,9307      & $00^h 40^m 48\fs 72$  & $-27^\circ 27\arcmin 07\farcs 2$
              & 1,704 & 0.10 & 0.0598 & 384 & 11.02 & 20 & 41 & 34 &  2 \\
MZ\,10167     & $01^h 51^m 12\fs 48$  & $-27^\circ 44\arcmin 09\farcs 6$ 
              & 1,599 & 0.14 & 0.0606 & 201 & 11.30 & 20 & 29 & 27 & 11 \\
MZ\,10300     & $02^h 24^m 27\fs 12$  & $-28^\circ 19\arcmin 19\farcs 2$
              & 1,624 & 0.15 & 0.0618 & 271 & 11.44 & 29 & 50 & 29 &  8\\
MZ\,10451     & $02^h 29^m 30\fs 48$  & $-29^\circ 37\arcmin 44\farcs 4$ 
              & 1,607 & 0.14 & 0.0610 & 455 & 11.64 & 40 & 75 & 44 & 18
\enddata
\tablecomments{Column~2+3: {\em GALEX} pointing coordinates.
  Column~4: effective FUV exposure time. Column~5: mean Galactic FUV
  extinction across the {\em GALEX} field. Column~6--8: mean group
  redshift, velocity dispersion, and total stellar mass of the group
  members within $R=1$~Mpc of the group center. Column~9: number of
  group members used to compute $\langle z\rangle$, $\sigma$, and
  $M_{\ast,\rm{tot}}$. Column~10: total number of spectroscopically
  identified group member candidates within the {\em GALEX}
  field. Column~11: number of group members within the {\em GALEX}
  field detected in both NUV and FUV bands. Column~12: number of group
  members within the {\em GALEX} field detected at 24\,$\mu$m.}
\end{deluxetable*}

For source identification, the {\sc SExtractor}--generated source
catalog provided by the {\em GALEX} pipeline was employed. Only
sources detected at S/N~$\ge 3$ in at least one of the two bands were
considered for further analysis. The adopted source positions are
based on an S/N--weighted average of the NUV and FUV positions. For
each group, the UV source catalog was cross-correlated with a list of
spectroscopically identified group members, flagging for each member
the nearest UV source within a maximum separation of $5''$ as a
match. This separation roughly corresponds to the full-width at half
maximum of the NUV point spread function (PSF). Table~\ref{tab,log}
lists for each group the number of spectroscopic (including candidate)
members $N_{\rm gal}$ within the {\em GALEX} field, and the subset of
those detected in both UV bands and at 24\,$\mu$m. Out of a total of
833 group members within the {\em GALEX} fields, 721 (622) are
detected in NUV (FUV), and 601 (72\%) are detected in both UV bands.

We focus here exclusively on the {\em integrated} UV properties of the
group galaxies, since the majority of the group members are only
slightly, if at all, resolved by {\em GALEX}. We adopted NUV and FUV
total magnitudes measured within elliptical apertures scaled to
2.5~times the Kron diameter, as provided by the {\em GALEX} pipeline
on the basis of sky-subtracted and response-corrected images. For
resolved sources, these apertures are generally comparable to the
$D_{25}$ aperture obtained from $B$-band photometry \citep{dona07}.

All magnitudes were corrected for Galactic extinction using the
reddening maps of \citet{schl98} and the results of \citet{wyde05}.
These corrections range from 0.1--0.5~mag, with a median extinction of
0.15~mag in both bands. Magnitudes were also $k$-corrected, by using
the {\sc galev} stellar evolutionary synthesis code \citep{kotu09} to
construct a grid of galaxy spectral models that cover the region
occupied by the group members in FUV--NUV vs.\ NUV--R color space. The
models assumed a Salpeter initial mass function (IMF) along with
various star formation histories and levels of intrinsic
extinction. For each observed group galaxy, predicted $k$-corrections
from the code associated with the nearest match in color space was
then applied. With the exception of very red early-types and strongly
star-bursting spirals, the corrections are generally $\lesssim
0.1$~mag in both bands for all group members. As mentioned, all UV
photometry for individual galaxies will be presented in a forthcoming
paper (L.~Bai et~al., in preparation).

With most of our group members remaining unresolved, we cannot
directly estimate the contribution of any active galactic nucleus
(AGN) activity to the UV flux. However, the fraction of AGN
identifiable on the basis of their optical emission lines or their
X-ray properties is small for the {\em XI} sample, at the few per~cent
level \citep{shen07,bai10}, so AGN contamination in the UV is
generally unlikely to be important.  In addition, the dependence of
the AGN fraction on global environment is generally inferred to be
weak \citep{arno09}, so we do not expect environmental variations in
AGN activity to significantly impact our results.

The approximate 100\% completeness limit for our {\em GALEX}
observations in either band can be estimated from the apparent
(i.e.\ not extinction-corrected) magnitude histograms of all sources
detected in the field of MZ\,770, the group with the shortest
identical exposure in FUV and NUV, and MZ\,4592, the group field
suffering the strongest Galactic extinction
(cf.\ Table~\ref{tab,log}). As can be seen from
Figure~\ref{fig,histo}, incompleteness in the NUV becomes apparent for
the MZ\,4592 field below $m_{\rm AB} \approx 22.0$. We therefore adopt
this value as our sample-wide UV completeness limit in the
following. Taking a more conservative limit, e.g., $m_{\rm AB} \approx
21.7$, would result in slightly poorer statistics without affecting
any of our conclusions. For comparison, \citet{hain11} estimate a 90\%
FUV completeness limit of $m_{\rm FUV}=22.5$ for {\em GALEX}
observations of similar duration and Galactic extinction.

\begin{figure}
\begin{center}
\epsscale{1.17}
\hspace{-2mm}
\plotone{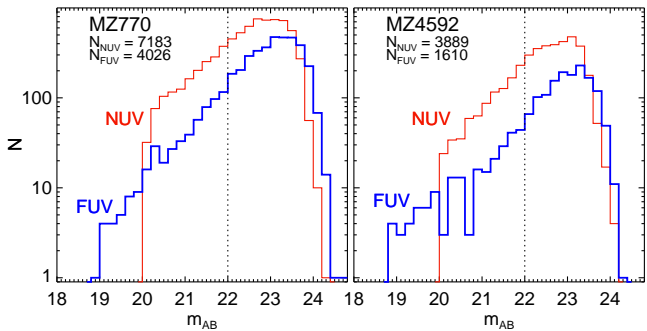}
\end{center}
\figcaption{Histograms of observed NUV and FUV magnitudes of all
  sources detected above $3\sigma$ significance in our shortest
  exposure, MZ\,770, and the one with highest Galactic extinction,
  MZ\,4592. Dotted vertical line outlines the approximate 100\%
  completeness limit adopted in this paper. Legends give the total
  number of NUV and FUV sources detected within each field.
  \label{fig,histo}}
\end{figure}

\subsection{Dust-corrected UV Star Formation Rates}

The UV light from star-forming galaxies is dominated by emission from
intermediate-mass stars younger than $\sim 10^8$~yr \citep{kenn98}.
Assuming that the SFR has remained roughly constant over this
timescale, it should be proportional to the intrinsic galaxy UV
luminosity $L_{\rm UV}$. Estimates of UV SFRs are model-dependent,
however, in part because they need to be corrected for attenuation by
dust (potentially by an order of magnitude), and in part because the
conversion of the dust-corrected $L_{\rm UV}$ to an SFR depends on the
assumed stellar metallicity and IMF, along with the geometry of the
distribution of dust and stars.

To estimate these dust corrections, we utilize our existing 24\,$\mu$m
photometry wherever possible and follow the approach of \citet{buat05}
to derive the FUV attenuation $A_{\rm FUV}$ as a function of FUV and
total infrared luminosity $L_{\rm TIR}$ (see \citealt{bai10}), using
\begin{equation}
A_{\rm FUV} = (-0.0333)y^3 + 0.3522y^2 + 1.1960y + 0.4967,
\label{eq,dust1}
\end{equation}
where $y= {\rm log} [L_{\rm TIR}/(\nu L_\nu)]$, $\nu =1.96 \times
10^{15}$~Hz is the effective frequency of the {\em GALEX} FUV band,
and $L_\nu$ is the measured specific FUV luminosity (in
erg~s$^{-1}$~Hz$^{-1}$). Dust-corrected star formation rates were then
obtained as \citep{sali07}
\begin{equation}
  \mbox{SFR$_{\rm FUV}$}\mbox{ (M$_\odot$~yr$^{-1}$)} = 1.08\times
  10^{-28}\,L_{\rm FUV},
\label{eq,SFR}
\end{equation}
where $L_{\rm FUV}$ is the dust-corrected specific FUV
luminosity. Equation~(\ref{eq,SFR}) is valid for a Salpeter IMF and a
mean stellar metallicity of $0.8Z_\odot$.

For the galaxies without a {\em Spitzer} detection, FUV magnitudes
were first corrected according to equation~(6) in \citet{sali07},
based on UV~magnitudes of normal (NUV--R$<4$) star-forming galaxies:
\begin{equation}
A_{\rm FUV} = \left\{\begin{array}{ll}
           2.99\, C_{\rm UV}+0.27, &  C_{\rm UV}<0.90 \\
	   2.96, &  C_{\rm UV}\geq 0.90, \end{array} \right.
\label{eq,dust2}
\end{equation}
where $C_{\rm UV}$~=~(FUV--NUV) is the rest-frame UV color.  The
associated dust-corrected SFRs were then compared to those derived on
the basis of equation~(\ref{eq,dust1}) for all galaxies detected at
both NUV, FUV, and 24\,$\mu$m. This comparison is shown in
Figure~\ref{fig,SFR}, as a function of the SFR from FUV$+$IR data
[i.e., equations~(\ref{eq,dust1}) and (\ref{eq,SFR})] and of galaxy
stellar mass. Using UV data alone on average tends to overestimate the
SFR relative to the result when combining UV and IR data, and this is
more prominent for galaxies with low SFR and high $M_\ast$ (see also,
e.g., \citealt{igle06,cort08a}).

\begin{figure}
\begin{center}
\epsscale{1.17}
\hspace{0mm}
\plotone{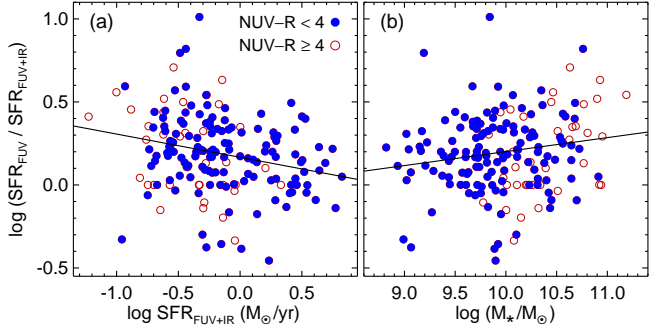}
\end{center}
\figcaption{SFR estimates with correction for dust from UV data only
  [equation~(\ref{eq,dust2})], relative to those obtained from a
  combination of UV and 24\,$\mu$m data [equation~(\ref{eq,dust1})],
  for the 169 galaxies detected at both NUV, FUV, and 24\,$\mu$m.
  Shaded circles show normal UV--blue star-forming galaxies, empty
  circles the remainder. Lines represent the best linear fits to the
  full data sets.
  \label{fig,SFR}}
\end{figure}

To bring our UV--only estimates into better agreement with those
obtained from the combined use of UV and 24\,$\mu$m data, we performed
a linear fit to the full data in Figure~\ref{fig,SFR}(b), implying
\begin{equation}
 \mbox{log}\,\left( \frac{\mbox{SFR}_{\rm FUV}}{\mbox{SFR}_{\rm
     FUV+IR}} \right) = 0.084\,\mbox{log}\,(M_\ast/M_\odot) - 0.64,
\label{eq,sfrcorr}
\end{equation}
with a mean absolute deviation of 0.18~dex (which is comparable to the
typical uncertainty on $M_\ast$). This result, shown as a solid line
in the Figure, is consistent with that obtained for the blue galaxies
only (NUV--R$<4$, for which equation~(\ref{eq,dust2}) is valid). We
use it to correct the inferred SFR for galaxies without a {\em
  Spitzer} detection at a given $M_\ast$. As will be discussed in
Section~\ref{sec,SFR}, the resulting SFRs are, on average, in very
good agreement with those obtained in other similar studies.  For our
adopted UV completeness limit and a minimum Galactic extinction of
0.1~mag (cf.\ Table~\ref{tab,log}), equation~(\ref{eq,SFR}) would
imply that we are complete to SFR$_{\rm FUV} <
0.06$~$M_\odot$~yr$^{-1}$ in all groups before correction for dust.

\subsection{Quantifying Galaxy Environment}\label{sec,environ}

The inferred UV and star formation properties of the group members are
here presented with a focus on their dependence on galaxy environment,
using total group stellar mass and/or velocity dispersion as main
indicators of global group environment. One further indicator,
discussed below, is obtained by dividing the groups into ones
containing UV--optical red and blue central galaxies, respectively.

Local galaxy environment is quantified by projected distance $R$ to the
adopted group center, optionally normalized by $R_{200}$. The latter
values were estimated for each group from the radial velocity
dispersion and mean redshift using
\begin{equation}
R_{200}=1.73\sigma [\Omega_\Lambda + \Omega_m(1+z)^3]^{-1/2}\mbox{ }
h^{-1}_{100}\mbox{ kpc}
\end{equation}
\citep{finn04}. Results range from $R_{200}\approx 150$--1200~kpc
(median of 520~kpc), implying that all groups are covered by {\em
  GALEX} to at least $2R_{200}$.

Another commonly used indicator of local environment is the galaxy
surface density $\Sigma_5 = 6/(\pi R_5^2)$, where $R_5$ is the
projected distance of each galaxy to its fifth nearest neighbor. This
can be reliably used across different systems, provided these are
globally broadly similar. However, our groups spans a wide range in
richness and estimated $R_{200}$, so at fixed 3--D galaxy density
$\rho_{\rm gal}$, projection effects would act to boost $\Sigma_5$
progressively more in the larger systems. To compensate for this, we
employ instead an estimate of $\rho_{\rm gal}$, derived by normalizing
$\Sigma_5$ by a characteristic radius $R^\prime$ for each
group. Noting that self-similarity would require $R^\prime \propto
\sigma \propto R_{200}$, we simply take $\rho_{\rm gal} = \Sigma_5
/R_{200}$. All results reported here are qualitatively similar though
if plotted against $\Sigma_5$ instead of $\rho_{\rm gal}$.

One further issue is our lower spectroscopic completeness outside the
$R\approx 15\arcmin \approx 1$~Mpc IMACS field-of-view, which could
act to systematically bias $\Sigma_5$ low at large radii. However, we
find no evidence for a break around $R\approx 15\arcmin$ in the
distribution of $\Sigma_5$ vs.\ $R$ for all group members.
Furthermore, the distributions of $M_R$ for the spectroscopically
identified members within and outside the IMACS field are broadly
similar down to at least $M_R\approx -19.0$, as shown in
Figure~\ref{fig,complete}. We find that all our results are
quantitatively similar, and all conclusions unaffected, if computing
$\Sigma_5$ by only including galaxies brighter than this
magnitude. Hence, to fully exploit our extensive IMACS spectroscopy,
which extends beyond $R_{200}$ for all groups and to $2R_{200}$ for
$\sim 75$\% of them, we compute $\Sigma_5$ and $\rho_{\rm gal}$ using
all available members.

\begin{figure}
\begin{center}
\epsscale{1.17}
\hspace{-2mm}
\plotone{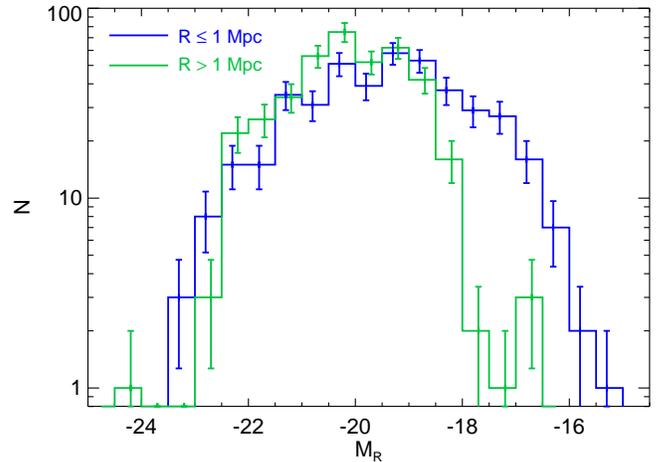}
\end{center}
\figcaption{Histograms of absolute $R$--band magnitudes for galaxies
  within the $R\approx 1$~Mpc IMACS field and for those covered by our
  auxiliary data at larger radii. Error bars represent Poisson errors.
\label{fig,complete}}
\end{figure}

\section{Results}

\subsection{Galaxy Colors}\label{sec,colors}

The UV--optical color, specifically NUV--R, is a useful diagnostic of
a galaxy's recent level of star formation \citep{sali07}. A
UV--optical color--magnitude diagram for all groups is
presented\footnote{As is customary, NUV magnitudes have here been
  corrected for Galactic extinction but not for any intrinsic
  reddening.} in Fig.~\ref{fig,colormag}, with each data point
color-coded according to its FUV--NUV color. The galaxy sample is seen
to split into a dominant population of star-forming blue-cloud members
joined by a smaller number of predominantly optically bright red
galaxies. Also shown is the region below the NUV sensitivity limit of
our data; this assumes a typical NUV $3\sigma$ limiting sensitivity of
$m_{\rm AB} \simeq 24.0$ after correction for Galactic extinction
(cf.\ Figure~\ref{fig,histo}).  This limit prevents the UV detection
of optically faint, passive galaxies, so Figure~\ref{fig,colormag}
alone may not provide a full picture of the extent of the passive
population in these groups. Incompleteness of the UV data will
generally be taken into account in the analysis which follows.

\begin{figure}
\begin{center}
\epsscale{1.17}
\hspace{0mm}
\plotone{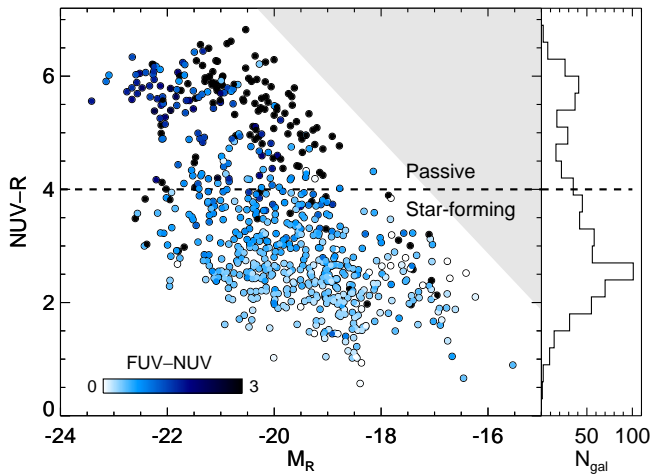}
\end{center}
\figcaption{UV--optical color--magnitude diagram of all individual
  UV--detected group members, color-coded according to FUV--NUV
  color. Dashed line at NUV--R\,=\,4.0 marks a commonly used
  distinction between star-forming and passive galaxies (e.g.,
  \citealt{sali07}). Shaded region represents the area below the
  sensitivity limit of our UV data. Right panel shows a histogram of
  the NUV--R colors.
   \label{fig,colormag}}
\end{figure}

An orthogonal regression fit to the NUV--R color as a function of
specific SFR (sSFR = SFR/$M_\ast$) shows that a value of
NUV--R\,$<$\,4 corresponds to SFR/$M_\ast > 10^{-10.5}$~yr$^{-1}$. As
will be shown in Section~\ref{sec,SFR}, considering galaxies above
this sSFR cleanly isolates the star-forming population that dominates
these groups. In the following, we will therefore use NUV--R\,$<$\,4
to separate "star-forming" galaxies from "passive" ones, noting that
this is also the limit to which equation~(\ref{eq,dust2}) remains
valid. Using this color cut, Figure~\ref{fig,colormag} confirms that
star-forming galaxies are generally relatively FUV bright, as
anticipated from equation~(\ref{eq,SFR}). In contrast, a large
fraction of the passive red galaxies have red FUV--NUV colors, with
almost half ($43\pm 5$\%) of them remaining undetected in the FUV.

Despite this, we note that bright, central early-type galaxies in
massive galaxy clusters {\em can} display blue colors and excess UV
light beyond that expected from old stars, indicating ongoing star
formation \citep{maha09,dona10,hick10}. This could potentially be
induced by cooling of the intracluster medium onto the massive galaxy
at the center of the cluster potential. As such, the color of the
central group galaxy might provide another indicator of the global
nature of the host environment. To examine this,
Figure~\ref{fig,BGGcol} shows the distribution of NUV--R colors for
the 22 central galaxies covered by {\em GALEX}, illustrating that the
majority (16) are clearly "passive", NUV--R\,$\ga 5$, and drawn from
the red-sequence population within the groups. Visual inspection of
our IMACS $R$-band images, taken in sub-arcsec seeing, confirms that
these 16 are all early-type galaxies.

\begin{figure}
\begin{center}
\epsscale{1.18}
\hspace{-2mm}
\plotone{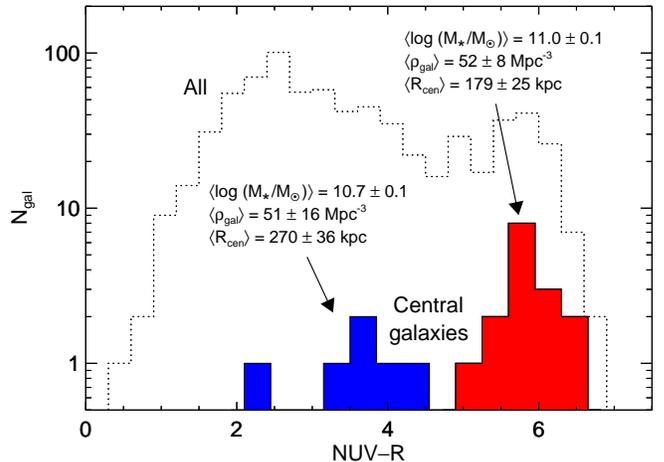}
\end{center}
\figcaption{Histograms of NUV--R colors for the central group galaxies
  (filled) and for all galaxies (dotted). Legends give the average
  $M_\ast$, $\rho_{\rm gal}$, and projected distance $R_{\rm cen}$
  from the group luminosity centroid, for the two classes of central
  galaxies.
 \label{fig,BGGcol}}
\end{figure}

However, a minority of the central galaxies have bluer colors, and all
but the reddest one of these are late-type galaxies. These could
potentially represent chance projections of blue galaxies situated at
large radii, rather than galaxies physically associated with the group
cores. They all have $M_R < -21.0$ and are situated $<0.4$~Mpc from
the group luminosity centroid. Based on the projected density of
equally bright, blue galaxies at larger radii, $R=1$--2~Mpc, we would
expect a total of only $0.9\pm 0.2$ such galaxies to be projected onto
this central region of the relevant groups. Thus, most of these six
blue centrals are likely to be physically situated in the group
core. On average, they have lower stellar masses than do the red,
passive ones, and they also reside further from the group luminosity
centroid and in groups of lower total stellar mass
($\langle$log\,$(M_{\ast,{\rm tot}}/M_\odot)\rangle = 11.28 \pm 0.09$
vs.\ $11.44 \pm 0.07$ within $R=1$~Mpc), velocity dispersion ($\langle
\sigma \rangle = 161 \pm 20$ vs.\ $282 \pm 28$~km~s$^{-1}$), and
richness ($\langle N_{\rm gal}\rangle = 26\pm 5$ vs.\ $42\pm 4$) than
do red central galaxies. This suggests that our blue, late-type
centrals reside in group halos that are relatively less massive and
dynamically younger, consistent with the finding that the fraction of
central group galaxies that are ellipticals is a strong function of
group halo mass \citep{wilm12}. Hence, these blue centrals likely
represent ordinary bright star-forming galaxies, rather than objects
with recent (dust-obscured) star formation fueled by cooling of
intragroup gas.

\subsection{Star-Forming Galaxy Fractions}\label{sec,uvfrac}

Given equation~(\ref{eq,SFR}), the fraction of group members detected
in the FUV is equivalent to the fraction of galaxies forming stars
above a certain rate. To explore how this quantity depends on galaxy
properties and environment, we first apply our cut at $m_{\rm AB} \leq
22.0$ in the FUV (i.e., SFR\,$\ga 0.05$\,$M_\odot$~yr$^{-1}$ before
dust correction) to homogenize UV completeness across the sample. The
resulting sample was also split at its median stellar mass of $M_\ast
= 10^{9.75}$\,$M_\odot$, to elucidate any dependence on galaxy
mass. Figure~\ref{fig,uvfrac} shows the resulting mean FUV fractions
as a function of local galaxy environment. The results reveal a clear
rise toward group outskirts, out to the largest radii and lowest
galaxy densities probed. This trend is robust in both galaxy stellar
mass bins when plotted against $\rho_{\rm gal}$, but is generally much
stronger for low-mass galaxies. These display lower UV fractions in
group cores than do high-mass ones, reversing the situation at large
radii where the higher UV fraction of low-mass galaxies simply
reflects the standard dependence of sSFR on galaxy mass (see
Section~\ref{sec,SFR}).

\begin{figure}
\begin{center}
\epsscale{1.17}
\hspace{-2mm}
\plotone{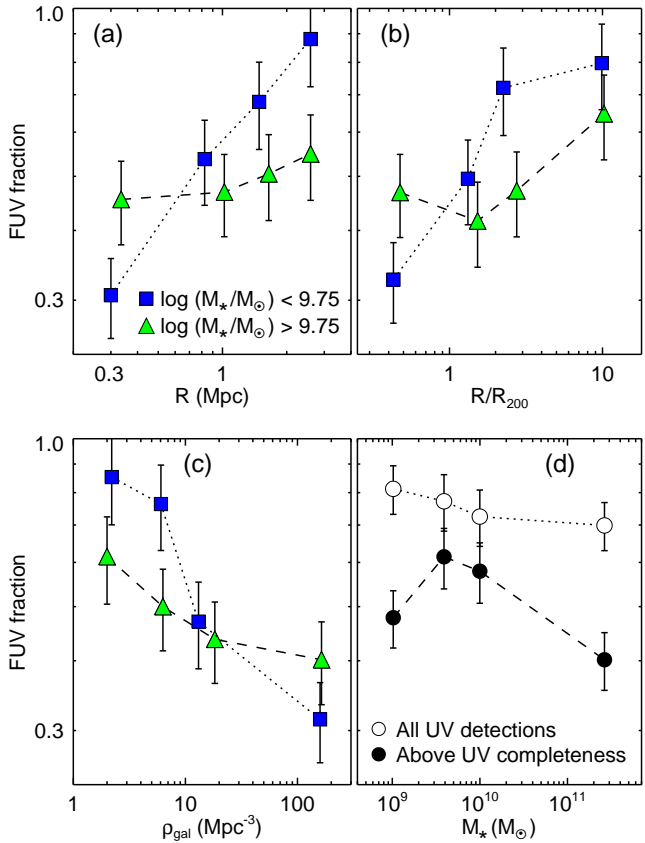}
\end{center}
\figcaption{Fraction of FUV detected group members above our UV
  completeness limit as a function of (a) projected radius from the
  group center, (b) radius normalized by $R_{200}$, (c) local galaxy
  density, and (d) galaxy stellar mass. In panels (a)--(c), squares
  and triangles show results for the low-- and high-mass half of the
  sample, respectively. Error bars represent Poisson errors on each
  bin.
\label{fig,uvfrac}}
\end{figure}

On average, the star-forming fraction among low-mass galaxies is
suppressed out to at least $R\approx 1.5$~Mpc and $R\approx 2R_{200}$,
thus mirroring results for much more massive clusters \citep{balo98}.
This implies that the group environment is influencing galaxy
properties out to similar overdensity radii as in much more massive
structures. If separating group and field/infall environments at these
radii, we find average star-forming (SF) fractions above the UV
completeness limit of $43\pm 3$\% in groups and $66\pm 6$\% in the
field. This implies a decrease in mean SF fraction by $35\pm 8$\% in
groups relative to the field, in agreement with the 30\% decline found
from our {\em Spitzer} data of nine of the groups \citep{bai10}.

These trends could potentially just reflect an underlying variation in
average galaxy stellar mass, with high-density group cores being
dominated by more massive, red, and generally UV-faint galaxies.
However, any residual mass dependence from bin to bin in panels
(a)--(c) is limited to a non-systematic variation in mean $M_\ast$
with environment of $<0.2$~dex. Furthermore, the decline in FUV
fraction with $M_\ast$ is weaker than that with environmental
parameters, as shown in Figure~\ref{fig,uvfrac}(d), even when
including the UV detections below our completeness limit. The
implication is that the observed trends are indeed environment--driven
rather than reflecting environmental variations in mean $M_\ast$.

To explore any dependence on global group environment,
Figure~\ref{fig,uvfrac2} displays the total FUV fractions within
$R_{200}$ as a function of total group stellar mass $M_{\ast,{\rm
    tot}}$, in four stellar mass bins containing a similar number of
groups. The data show a systematic decline in average star-forming
galaxy fraction with group stellar mass and velocity dispersion, from
$\sim 70$\% in the lowest-mass bin to $\sim 30$\% for the most massive
systems. Although subject to larger errors, the overall trend persists
if subdividing the galaxy population into a low- and high-mass half
(see figure inset), and so is not driven by systematic differences in
average galaxy $M_\ast$ with total group stellar mass. Hence, even
across the range of relatively low host halo masses considered here,
the overall star formation activity is systematically modulated by
global environment.

\begin{figure}
\begin{center}
\epsscale{1.17}
\hspace{0mm}
\plotone{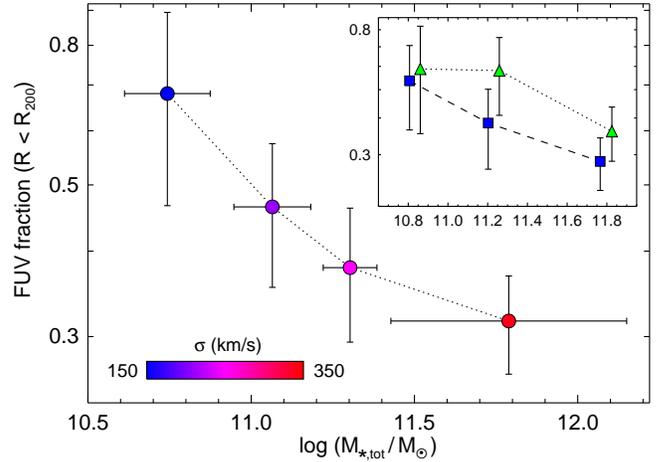}
\end{center}
\figcaption{Fraction of FUV detected group members above our UV
  completeness limit as a function of total stellar group mass (both
  evaluated within $R_{200}$). Vertical error bars represent Poisson
  errors on each bin, and color-coding illustrates the mean velocity
  dispersion within each group stellar mass bin. The partially covered
  MZ\,770 is not included. Inset shows the results (same axis units)
  when splitting the galaxy sample at the median stellar mass; symbols
  are as in Figure~\ref{fig,uvfrac}.
   \label{fig,uvfrac2}}
\end{figure}

\subsection{Star Formation Rates}\label{sec,SFR}

For comparing dust-corrected SFRs of the 601~galaxies detected in both
UV bands, we bring all objects on an equal footing by presenting their
specific SFR in Figure~\ref{fig,SSFR}. 
\begin{figure*}
\begin{center}
\epsscale{0.85}
\hspace{0mm}
\plotone{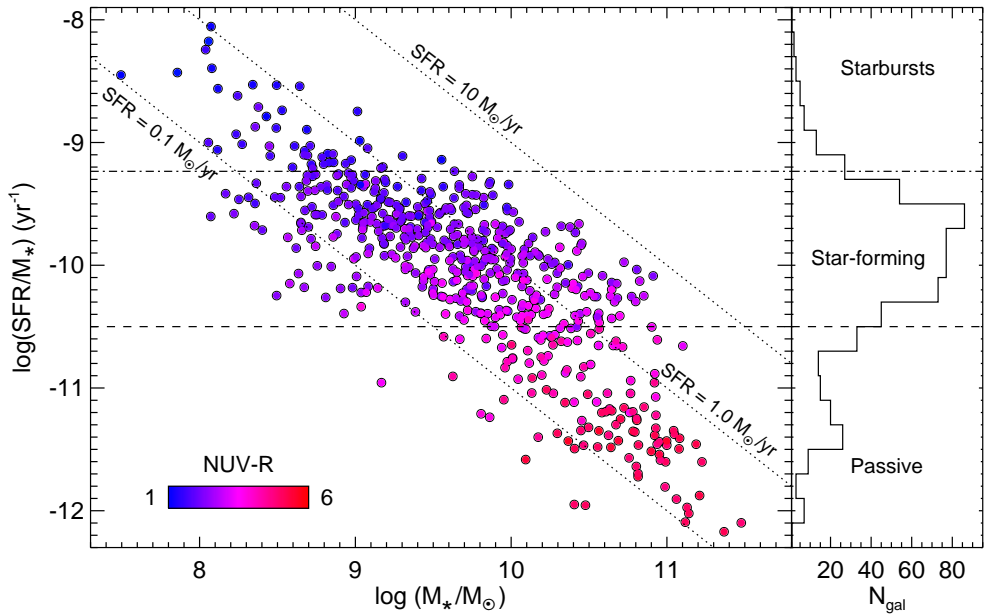}
\end{center}
\figcaption{Dust-corrected specific SFRs as a function of galaxy
  stellar mass. Galaxies are color-coded according to their
  UV--optical color, with dotted lines marking characteristic constant
  SFRs. Dashed line corresponds to an NUV--R color of 4.0 (see
  Section~\ref{sec,colors}), blueward of which galaxies are assumed to
  be star-forming. Galaxies above the dot--dashed line are currently
  forming stars at $>3$ times their past average, assuming star
  formation was initiated at $z=3$. Histogram shows the distribution
  of specific SFRs.
  \label{fig,SSFR}}
\end{figure*}
Above our approximate completeness limit, there is a clear decline in
the average sSFR with $M_\ast$ across three orders of magnitude in
stellar mass, and a corresponding increase in the UV--optical
color. The dashed line in the Figure shows the value of sSFR =
$10^{-10.5}$~yr$^{-1}$ that corresponds to the adopted distinction
between passive and star-forming galaxies at NUV--R\,=\,4, as
suggested by a regression fit between the two quantities; this is a
reliable and robust criterion, with only 3\% of the bluer group
members having estimated sSFRs below this value. The distribution of
sSFR values also mirrors the bimodality seen for the NUV--R colors,
and both distributions are clearly continuous, with no gap at
intermediate colors or sSFRs.

For the star-forming galaxies with log\,$(M_\ast / M_\odot ) > 8.5$,
an outlier-resistant least-squares fit\footnote{Using the {\sc ladfit}
  routine in the Interactive Data Language ({\sc idl})} to the data in
Figure~\ref{fig,SSFR} suggests a scaling relation of the form
\begin{equation}
  \mbox{log (SFR}/M_\ast) = -0.46\,\mbox{log}\,(M_{\ast}/M_\odot)-5.4,
\label{eq,ssfr}
\end{equation}
with a mean absolute deviation of 0.24~dex. The result implies SFR\,$
\propto M_{\ast}^{0.54}$. This is nominally slightly flatter than, but
broadly consistent with, the relation obtained from {\em GALEX} data
of much larger samples of the global low-redshift galaxy population,
SFR\,$\propto M_{\ast}^{0.64-0.65}$ \citep{sali07,schi07}. Some
flattening of this relation in groups relative to the field may in
fact be expected, if the SFR of low-mass galaxies is suppressed in
groups. Indeed, at the median log\,$(M_\ast / M_\odot ) = 9.63$ of the
relevant galaxies, equation~(\ref{eq,ssfr}) would predict a mean sSFR
which is 38--45\% lower than seen for the field in the above studies
(when corrected for differences in the assumed IMF, where
necessary). As will be shown, this difference is significant and in
excellent agreement with that inferred from the {\em XI} data
themselves. The important implications are that the mean sSFR of
star-forming galaxies is suppressed in groups relative to the field at
a given $M_\ast$, and that our estimated SFRs are not significantly
biased compared to those derived in other similar studies.

To further explore this suppression in sSFR, we plot in
Figure~\ref{fig,SSFR_rad} the sSFR of star-forming galaxies against
local environment in equal-number bins. The sample has further bin
equally split into three bins in stellar mass, to test for systematic
dependencies for low-- and high--mass galaxies separately. Any
residual mass dependence {\em within} these bins is negligible, as the
mean $M_\ast$ within each bin varies non-systematically with
environment by $\la 0.1$~dex in all three diagrams. Environmental
trends are clearly present, particularly for the lowest-mass galaxies,
where the mean sSFR declines by 0.4~dex across the region probed. At
$M_\ast \ga 1\times 10^{10} M_\odot$, any systematic trends become
negligible, however. Hence, not only does the fraction of SF galaxies
vary with environment for our sample, but the actual SFR {\em within}
star-forming galaxies does so too at fixed galaxy stellar mass $M_\ast
\la 10^{10} M_\odot$. Note that, except for the highest--$M_\ast$ bin,
the suppression of SFR/$M_\ast$ is significant out to at least $R\sim
1.5$~Mpc~$\approx 2R_{200}$, similar to what is seen for the SF
fractions themselves.

\begin{figure*}
\begin{center}
\epsscale{0.95} 
\plotone{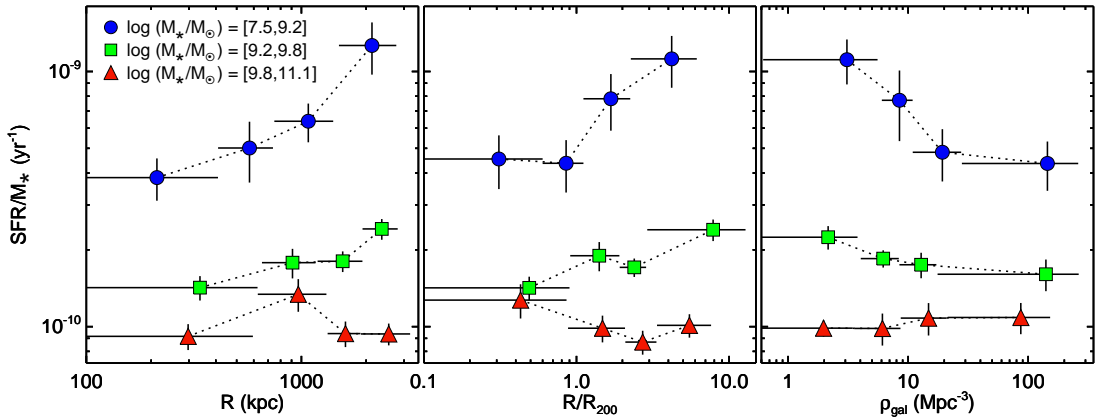}
\end{center}
\figcaption{Mean specific SFRs of star-forming galaxies (with
  NUV--R~$<4$) as a function of local galaxy environment, in three
  bins of galaxy stellar mass. Vertical error bars represent errors on
 the mean in each bin.
   \label{fig,SSFR_rad}}
\end{figure*}

To our knowledge, this is the first time such a trend has been
reported. No such trends were noticed in our earlier study (when
considering galaxies at $M_\ast > 10^{9.6} M_\odot$; \citealt{bai10}),
plausibly due to poorer statistics, smaller spatial coverage, and the
fact that the effect is subtle for galaxies above this mass limit.
The result suggests that with the full {\em XI} sample, we are now for
the first time witnessing the {\em ongoing} quenching of star
formation within nearby galaxy groups. Hence, such quenching is still
taking place in the local universe within the relatively low-mass
galaxy structures that dominate galaxy redshift surveys.

\subsection{Fraction of Starburst Galaxies}

In massive galaxy clusters, infalling galaxies can show enhanced star
formation at intermediate radii (e.g., \citealt{port08,pere10}),
possibly due to interactions with other nearby galaxies.
Figure~\ref{fig,SSFR_rad} suggests that a similar effect does not
occur in smaller groups. To verify this, starburst galaxies were first
identified within the groups. In line with our previous analysis of
the 24\,$\mu$m data \citep{bai10}, starbursts are here defined as
galaxies with a current SFR of at least three times their past
average, i.e.\ with birthrate parameters $b>3$, where
\begin{equation}
   b= \frac{{\mbox{SFR}}(z=0.06)}{\langle {\mbox{SFR}}\rangle _{\rm
       past}} = \frac{\mbox{SFR}\times \tau}{M_\ast}(1-f_R),
\label{eq,sb}
\end{equation}
and where $\tau$ and $f_R$ is the typical stellar age and gas
recycling fraction, respectively. Assuming $f_R=0.5$ and $\tau$
corresponding to a formation redshift $z=3$, just prior to the peak in
the cosmic SFR density, the resulting limit of $b\geq 3$ is shown as a
dot--dashed horizontal line in Figure~\ref{fig,SSFR}. 

The spatial variation in starburst galaxy fraction is illustrated in
Figure~\ref{fig,starbursts}. This plot confirms the absence of a
characteristic density or radius at which star formation is enhanced,
with fractions declining, within the errors, monotonically and
significantly toward group cores. Within the uncertainties, there is
also no systematic dependence of the starburst fraction within
0--$R_{200}$, 1--2$R_{200}$, or 2--3$R_{200}$ on total group stellar
mass. Hence, the enhanced starburst activity reported around galaxy
clusters does not persist into the group regime, not even for the most
massive groups in the sample. The unlikely alternative is that any
localized enhancement in sSFR occurs at $R\ga 2.5$~Mpc,
i.e.\ partially beyond the $R\approx 2$-3~Mpc region identified for
clusters \citep{port08}.

\begin{figure}
\begin{center}
\epsscale{1.17}
\hspace{0mm}
\plotone{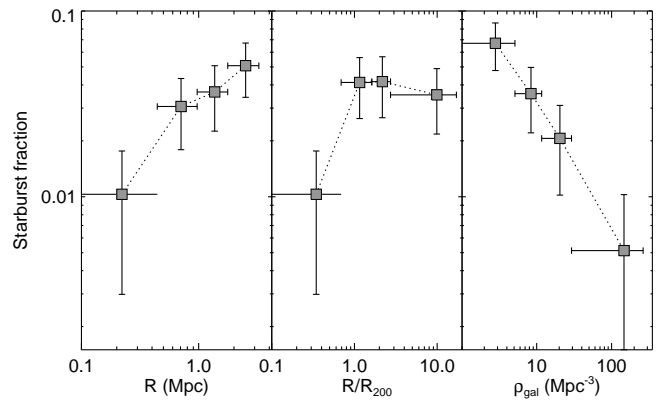}
\end{center}
\figcaption{Fraction of galaxies with $M_\ast > 10^{8.5} M_\odot$ that
  are starbursts above the UV completeness limit as a function of
  local galaxy environment. Vertical error bars represent Poisson
  errors on each bin.
   \label{fig,starbursts}}
   \end{figure}

Among the group members with $M_\ast > 10^{8.5} M_\odot$ (for which
our detection of starburst galaxies should be complete), a small
fraction $f_{\rm SB} = 25/778$ ($3.2\pm 0.7$\%) can be classified as
starbursts above our UV completeness limit. This is slightly higher
than the value of $f_{\rm SB} < 1$\% found for similar-mass galaxies
in our 24\,$\mu$m data of a subset of the group sample. As
demonstrated in Figure~\ref{fig,starbursts}, the difference is at
least partly explicable on account of the larger physical region
probed here; for example, the fraction is reduced to $2.4\pm 0.7$\%
within $2R_{200}$ and to $1.0\pm 0.6$\% within $R=0.7$~kpc, the region
probed by our {\em Spitzer} MIPS data in the narrowest direction. The
inferred fraction within the full {\em GALEX} field is also consistent
with the value of $f_{\rm SB}= 7.6^{+12.6}_{-5.4}$\% derived for the
local universe by \citet{sarg12}. The absence of starbursts with
$M_\ast > 1\times 10^{10} M_\odot$ in Figure~\ref{fig,starbursts} is
furthermore consistent with the lack of enhanced star formation seen
within groups at $z \sim 0.4$ down to this mass completeness limit
\citep{balo09}.

\section{Discussion}\label{sec,discuss}

The inference that SFR/$M_\ast$ within star-forming galaxies is
environmentally suppressed in groups relative to the field is a novel
result, to the best of our knowledge. It suggests that we are directly
observing the ongoing quenching of star formation in dense
environments. We speculate that this result was not identified in
previous group studies for at least three reasons. First, while our
group sample is homogeneously selected, the global group properties
are very heterogeneous, with groups spanning at least an order of
magnitude in velocity dispersion, total stellar mass, richness, and
X-ray luminosity (see Table~\ref{tab,log} and
\citealt{rasm06b,rasm10}).  Combined with the large {\em GALEX} field
which covers galaxies well into the infall regions, this allows us to
self-consistently probe the full range of local and global galaxy
environments up to the scale of massive groups. Second, many of these
groups are likely dynamically young \citep{bai10} and so contain a
significant population of gas-rich star-forming galaxies which have
been only mildly, if at all, affected by the group environment. If
much of the environmentally induced gas loss from galaxies, e.g., via
starvation, occurs when the group or cluster first collapses
\citep{lars80}, then the inclusion of such dynamically unevolved
systems is crucial for observing SF quenching in action. Finally, our
deep spectroscopy enables inclusion of numerous low-mass galaxies down
to $M_\ast \ga 10^8 M_\odot$, for which environmental effects are
found to be be particularly prominent. This is only feasible for
nearby group samples such as this; other systematic multi-wavelength
studies of groups focus instead on systems at intermediate redshifts
(e.g., \citealt{wilm05,mcge11}) for which the limiting galaxy stellar
mass is significantly higher.

\subsection{Star Formation and Galaxy Environment}

Our results show that both the star-forming and starbursting galaxy
fractions decline toward dense group cores, in general agreement with
results for other group samples at low and intermediate redshifts
\citep{balo04,wilm05,jelt07,mcge11} and with those from our 24\,$\mu$m
imaging \citep{bai10}. Importantly, the decline in SF fraction is
present also at fixed stellar mass, and with galaxies below $M_\ast
=10^{9.5-10}$~$M_\odot$ showing stronger environmental trends than do
more massive objects.

Globally, the SF fractions within $2R_{200}$ are 35\% lower than seen
at the largest radii probed. Comparing this to our corresponding
24\,$\mu$m results for a subset of this sample \citep{bai10} suggests
excellent agreement between the UV and mid-IR results, but definitive
confirmation of this will have to await the {\em Spitzer} results for
the full sample (L.~Bai et~al., in preparation). In contrast, the
observed systematic decline in SF fraction with total group stellar
mass was not seen in our preliminary {\em Spitzer} analysis, possibly
due to poorer statistics and the consideration of a fixed metric
region for all groups.

For star-forming galaxies alone, a systematic suppression of sSFR with
local environmental parameters is further seen, persisting out to the
largest radii and lowest galaxy densities probed. This is
preferentially observed within low-mass galaxies, with the trend
becoming negligible at stellar masses $M_\ast \ga 1\times
10^{10}$~$M_\odot$. At fixed, low galaxy stellar mass, the suppression
is also modest, amounting to a factor of $\la 2.5$ for SF galaxies in
the densest group regions (Figure~\ref{fig,SSFR_rad}). This is much
less than the corresponding variation with stellar mass for the same
galaxies ($\sim 2$~orders of magnitude; cf.\ Figure~\ref{fig,SSFR}).
Nevertheless, if taking galaxies at $R\ga 1.5$~Mpc\,$\approx 2R_{200}$
as representative of the field and infall regions (see
Figures~\ref{fig,uvfrac} and \ref{fig,SSFR_rad}), then the mean and
$1\sigma$ error on the sSFR of all SF galaxies in these regions is
$4.2\pm 0.8\times 10^{-10}$~yr$^{-1}$, significantly higher than the
$2.8\pm 0.3\times 10^{-10}$~yr$^{-1}$ found within the group
environment itself. Given that the mean stellar mass of the two
sub-populations are similar, this implies an average suppression in
star formation activity {\em within} all star-forming galaxies in
groups by $34\pm 14$\% relative to the ``field'' value within our
data. This is in good agreement with the observed $\sim 40$\% decline
relative to typical values for large local galaxy samples discussed in
Section~\ref{sec,SFR}.

The observed environmental variation in sSFR is at odds with the
generally negligible difference between typical sSFRs in groups and
the field seen in other studies of group samples at low and
intermediate redshifts \citep{balo04,vulc10,mcge11}. The inclusion of
galaxies well below the $M_\ast \ga 3\times 10^9 M_\odot$ limit of
these other studies ---afforded by the proximity of our sample--- as
well as the nature of the {\em XI} groups and of the present data (as
described above) may all contribute to this difference.

Before interpreting these results, we note that two effects could bias
the results in Figures~\ref{fig,uvfrac} and \ref{fig,SSFR_rad}. First,
the differences in spectroscopic completeness between the $R\approx
1$~Mpc IMACS data and the auxiliary data employed at larger radii
imply a relative shortfall of low-mass (i.e., high-sSFR) galaxies at
$R \ga 1$~Mpc in the sample. However, this would only act to suppress
the observed trends, which could therefore be even stronger than
inferred. Second, the presence of interlopers, unaffected by the group
environment, would also dampen any real trends. The importance of this
can be assessed by quantifying the projected density of SF background
galaxies as determined at large radii. To do so, we take galaxies at
and beyond the outermost data points in Figures~\ref{fig,uvfrac} and
\ref{fig,SSFR_rad}, ie.\ those with $\rho_{\rm gal} < 3$~Mpc$^{-3}$ or
$R/R_{200} > 5$, as belonging to the field. In addition, we only
consider galaxies with $M_R< -19$, in order to homogenize
spectroscopic completeness across all radii. This yields a mean
$\Sigma_5 = 0.7 \pm 0.1$~Mpc$^{-2}$ for the projected density of such
SF galaxies in the field. Within the group region ($R \la 2R_{200}$),
the corresponding value is $9.6\pm 1.3$~Mpc$^{-2}$, rising to $\approx
15$~Mpc$^{-2}$ within $R_{200}$. These results do not change if only
considering galaxies above our UV completeness limit, suggesting that
no more than 5--7\% of the $M_R < -19$ SF galaxies seen within our
groups are interlopers.

For fainter galaxies, a larger interloper fraction cannot be ruled
out, due to the lower spectroscopic completeness at large
radii. However, we do note that the 5--7\% would represent a {\em
  global} upper limit if the fraction of (low-mass, high-sSFR)
galaxies fainter than $M_R= -19$ is lower in groups than in the
field. This assertion would be in line with the observed decline in
the dwarf-to-giant galaxy ratio toward cluster cores \citep{sanc08}.
The tentative implication is that the great majority of SF galaxies
identified within the {\em XI} groups are physically associated with
the group, even those in the densest regions of our most massive
groups.  In any case, a larger interloper fraction would only act to
suppress the observed trends, which are most prominent for the
lowest-mass galaxies, so in that sense our results are conservative.

In summary, the dependence of both SF fractions and the sSFRs of SF
galaxies on local group environment is a robust result. While the SFRs
depend primarily on stellar mass above our UV completeness limit, with
SFR\,$\propto M_\ast^{\sim 0.5}$, there is a significant residual
dependence on local environment at fixed stellar mass. Local galaxy
environment thus plays an important role in regulating both the
proportion and actual activity of star-forming galaxies in groups. The
fact that the SF galaxy fraction generally declines with increasing
total stellar group mass suggests that {\em global} group environment
may also modulate star formation. To explore this possibility, we plot
the star-forming fractions and the sSFR of star-forming galaxies as a
function of $\rho_{\rm gal}$ in Figure~\ref{fig,environ} while
controlling for global group parameters. 
\begin{figure*}
\begin{center}
\epsscale{0.98}
\hspace{-2mm}
\plotone{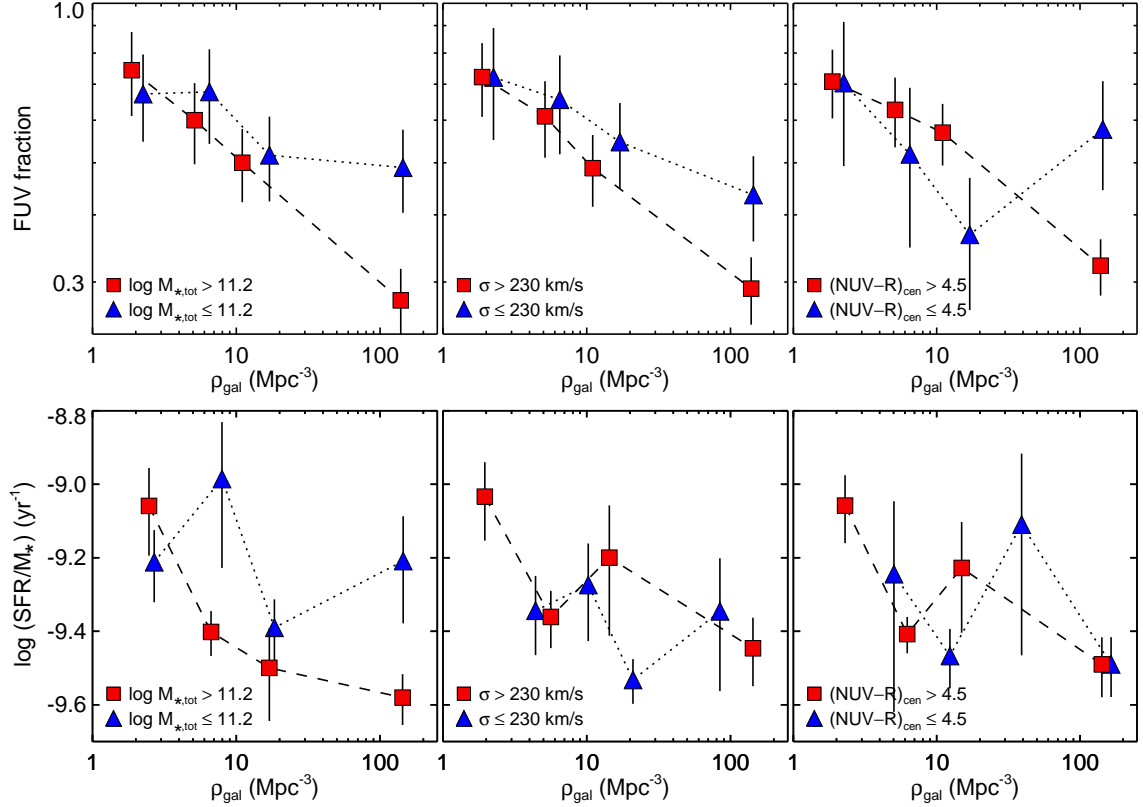}
\end{center}
\figcaption{Top panel: Fraction of FUV--detected galaxies above the UV
  completeness limit as a function of $\rho_{\rm gal}$ in equal-number
  bins. Sample is divided according to the host group median
  $M_{\ast,\rm{tot}}/M_\odot$, median $\sigma$, and color of the
  central group galaxy, respectively. Error bars represent Poisson
  errors. Bottom panel: Mean specific SFR of the low-mass ($M_\ast <
  10^{9.6} M_\odot$) star-forming galaxies in equal-number bins, with
  the sample divided as above. Error bars represent errors on the
  mean.
\label{fig,environ}}
\end{figure*}
To maximize the strength of any trends, the bottom panel only
considers the low-mass half of the star-forming galaxies, for which
the dependence of sSFR on local environment is strongest. In this
panel, the bin-to-bin variation in mean $M_\ast$ with environment is
$\la 0.2$~dex and not systematically dependent on $\rho_{\rm gal}$ or
$\Sigma_5$. The results indicate that, at fixed $\rho_{\rm gal}$, the
star-forming galaxy fraction and the sSFR of star-forming galaxies are
more heavily suppressed in the cores of high-mass groups with
``evolved'' central galaxies.

In fact, the trends in Figure~\ref{fig,SSFR_rad} are mainly driven by
galaxies in the latter groups, since no clear environmental
suppression of sSFR is seen for the low-mass groups with blue central
galaxies. To understand whether the absence of a trend in the low-mass
groups is driven simply by the arguably less evolved systems with blue
centrals, we subdivide the low-mass groups according to central galaxy
color. Figure~\ref{fig,lowmass} shows the resulting mean sSFRs as a
function of environment for these two classes of groups. We focus
again on the low-mass galaxies only ($M_\ast < 10^{9.6}$\,$M_\odot$),
in order to maximize the strength of any trends. While the low-mass
groups with red centrals do show mild evidence of a trend with $R$,
the mean specific SFRs of galaxies within the two types of groups are
similar within $R \sim 1.5$~Mpc~$\sim 2R_{200}$. Hence within the
relevant radii and densities, there is no strong evidence for ``less
evolved'' systems of a given low $M_{\ast,{\rm tot}}$ to behave
dramatically different from those containing red centrals.

\begin{figure}
\begin{center}
\epsscale{1.17}
%\hspace{-4mm}
\plotone{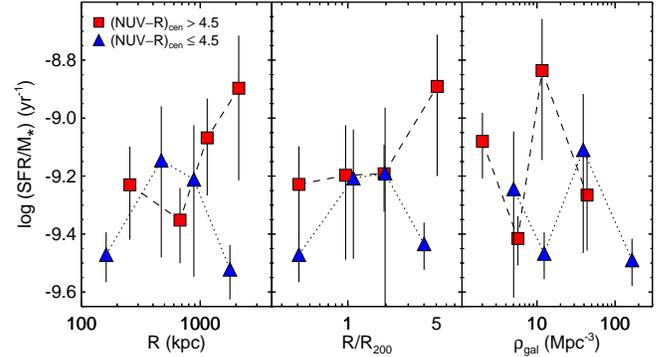}
\end{center}
\figcaption{Mean specific SFR of low-mass ($M_\ast <
  10^{9.6}$\,$M_\odot$) star-forming galaxies in the low-mass groups
  ($M_{\ast,\rm{tot}} < 10^{11.2}$\,$M_\odot$), subdivided according
  to central galaxy color.
\label{fig,lowmass}}
\end{figure}

The above results suggest that, at fixed local galaxy environment, a
residual contribution to quenching is present whose impact scales with
total group stellar mass. One caveat to this interpretation is that
$\rho_{\rm gal}$ (and $\Sigma_5$) may measure different "environments"
in small and large systems, with a given value of $\rho_{\rm gal}$
being representative of a relatively larger range in $R/R_{200}$ in
small systems. However, all trends in Figure~\ref{fig,environ} are
qualitatively similar, and all conclusions unaffected, if instead
plotting the results against $R$ or $R/R_{200}$. To further verify
that the inferred variations in SF activity at fixed (high) $\rho_{\rm
  gal}$ are not just reflecting variations in $\rho_{\rm gal}$ itself
with total group stellar mass, we divide the star-forming galaxies
according to their median $\rho_{\rm gal}$, consider only galaxies
within $R_{200}$, and control for galaxy stellar mass by only
considering the low-mass half of the resulting sample. The average
sSFR of these galaxies in the two bins of $\rho_{\rm gal}$ are shown
as a function of $M_{\rm *,tot}$ in Figure~\ref{fig,massdepend}.  A
decline with $M_{\rm *,tot}$ is seen in both bins of $\rho_{\rm gal}$,
confirming that the suppression in SF at fixed local environment does
scale with total group stellar mass.

\begin{figure}
\begin{center}
\epsscale{1.17}
\hspace{0mm}
\plotone{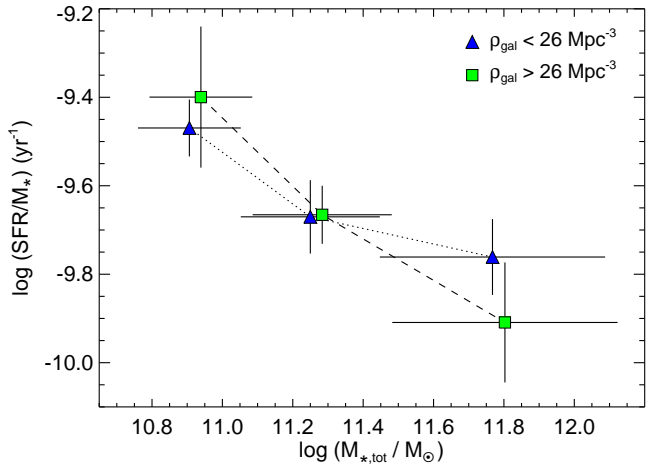}
\end{center}
\figcaption{Average specific SFR of low-mass ($M_\ast <
  10^{9.6}$~$M_\odot$) star-forming galaxies within $R_{200}$ as a
  function of total group stellar mass. Sample is divided according to
  local galaxy density. Results for the high-$\rho_{\rm gal}$ bin have
  slightly displaced along the $x$-axis for clarity.
\label{fig,massdepend}}
\end{figure}

\subsection{Quenching of Star Formation in Groups}

The immediate interpretation of the above results is that the
suppression of SF depends on both local and global group environment,
and so is possibly modulated by at least two different
mechanisms. Galaxy density, or some quantity scaling with it, plays a
role in shutting down SF in groups. Tidal interactions and mergers are
plausible candidates, as these should be important in these
low--$\sigma$ environments, especially in regions of high galaxy
density. In addition, quenching is more efficient at fixed galaxy
density in higher-$\sigma$ groups, opposite to what is expected from
tidal galaxy--galaxy interactions. If the latter are important, then
an additional mechanism whose efficiency is independent of $\rho_{\rm
  gal}$ but instead scales with group stellar mass and acts
predominantly in the densest regions must also have an impact.
Gas-dynamical processes such as ram pressure/viscous stripping and
starvation are obvious candidates.

Possible clues to the quenching mechanism emerge from the result that
an environmental dependence of sSFR in blue, star-forming galaxies at
fixed (low) galaxy mass is mainly observed in the more massive groups
with red central galaxies, not within the lowest-mass ones with blue
centrals. Nor is such a trend observed in massive low-$z$ clusters
\citep{wolf09,chun11,lu12}, although some evidence is seen when
including SF galaxies in $\sigma < 500$~km~s$^{-1}$ groups down to
sSFR\,$\approx 10^{-11}$~yr$^{-1}$ \citep{vond10}. To explain these
results, one could hypothesize that the blue SF galaxies observed in
low-$z$ clusters, which comprise a significantly smaller fraction than
in groups (cf.\ \citealt{bai10}), are strongly dominated by
interlopers associated with the general field (as supported by the
inference that backsplash galaxies in clusters have essentially no SF
left after just a single passage through the cluster;
\citealt{maha11}). Hence, little difference in sSFR between blue field
and ``cluster'' SF galaxies should be expected. On the other hand, in
the smallest and youngest groups (some of which are perhaps collapsing
only now) most SF galaxies will be real group members in which star
formation may not yet have been strongly suppressed. Only in
relatively massive, evolved groups below the scale of clusters can we
directly observe the ongoing environmental quenching of SF.

This raises the possibility of constraining the timescales $t_{\rm q}$
for quenching SF in groups. In order to see a correlation between sSFR
and clustercentric radius one needs a coincidence between $t_{\rm q}$
and the crossing timescale $t_{\rm cr}$. Conversely, if $t_{\rm q}\ll
t_{\rm cr}$ (``case~A'') then all SF will be shut down near the virial
radius, so there will be little radial trend further in (where SF
galaxies will mainly be interlopers, and the radial trend in SF
fraction will reflect that of the interloper fraction). In contrast,
if $t_{\rm q} \gg t_{\rm cr}$ (``case~B'') then galaxies will perform
multiple crossings as SF is quenched, so again there will be little
radial trend. For virialized systems which have collapsed at the same
epoch, $t_{\rm cr}$ is independent of system mass, since $t_{\rm cr}
\propto R/\sigma$ and $\sigma \propto (M/R)^{1/2} \propto R$. Hence,
the interpretation suggested here is that the quenching timescale
$t_{\rm q}$ is generally shorter in more massive systems, with rich
clusters falling under case~A above, the poorest (and/or youngest)
groups under case~B, and with relatively massive groups ($\sigma \sim
250$--500~km~s$^{-1}$) as the only systems in between.

For the $M_{\ast,{\rm tot}} > 10^{11.2}$\,$M_\odot$ half of our group
sample, the mean linear crossing time \citep{rood78} is in the range
$t_{\rm cr} = 1.1$--1.6~Gyr, depending on the maximum radius
considered ($R_{200}$, 2$R_{200}$, or $R=1$~Mpc). Since SF is not
completely shut down after a single crossing, this would suggest a
typical quenching timescale of $t_{\rm q} \ga 2$~Gyr in these
systems. This result provides strong independent confirmation of
recent estimates of $t_{\rm q} \ga 2$--3~Gyr based on simple models
for the accretion of galaxies onto groups \citep{mcge09,mcge11}.

If galaxy--galaxy interactions alone induce quenching through rapid
consumption of gas associated with an interaction-triggered starburst
phase, then the presently observed starburst fraction of $f_{\rm sb}
\approx 0.02$ inside $R\approx 2R_{200}$ limits the characteristic
timescale of this phase \citep{bai10}. The SF fraction in groups has
declined relative to that of the field since at least $z \approx
0.5$--0.6 \citep{wilm05,mcge11}, i.e., over the past $\sim 5$~Gyr. If
mergers and interactions have accomplished this with a constant
$f_{\rm sb}$ with redshift, then a quenched fraction of $f_{\rm q}
\approx 0.35$ today requires a typical quenching timescale $\Delta
t_{\rm q} \approx 5 \mbox{ Gyr}\times f_{\rm sb}/f_{\rm q} \la
0.3$~Gyr. This timescale is too short in comparison to the crossing
times mentioned above. In addition, there is no indication of enhanced
SF or starburst activity at small-to-intermediate radii, which could
otherwise indicate a temporary boost in SF associated with, or
immediately prior to, the actual quenching.

In this context, it is interesting to note the detection of a
pronounced bimodality in sSFRs in Hickson Compact Groups (HCGs;
\citealt{tzan10,walk12}), with a sparsely populated region separating
low-- and high-sSFR systems ($<10^{-10.5}$~yr$^{-1}$ and
$>10^{-9.9}$~yr$^{-1}$ respectively; see also Figure~\ref{fig,SSFR}).
This feature is more prominent than seen for our groups, other groups,
in the field, or in cluster cores \citep{walk12,wetz12}. It favors a
rapid transition from SF to quiescence, perhaps preceded by a period
of enhanced SF activity in at least some galaxies. The lack of such a
clear bimodality within the {\em XI} sample may indicate a slower
transition from SF to quiescence than in HCGs, or that SF is not
similarly enhanced prior to the transition. The latter possibility may
help to explain the lack of a clear enhancement in average sSFR or
starburst fraction at intermediate radii for our SF population. Since
galaxy--galaxy interactions should be particularly efficient in HCGs,
this further supports the idea that other quenching mechanisms are
playing a relatively more important role in the {\em XI} groups.

In summary, the results inferred from Figure~\ref{fig,environ}, the
required quenching timescales in comparison to that associated with an
interaction--induced starburst phase, and the comparison to HCGs, all
point to galaxy--galaxy interactions being complemented by a more
slowly acting quenching mechanism within our groups. Ram pressure
stripping and starvation are both likely to act over Gyr timescales in
these relatively low-mass environments. The latter process seems
particularly promising, as it can still remove hot galactic halo gas
even within fairly small groups \citep{kawa08}. Although starvation
alone is not expected to act efficiently out to such large radii and
low galaxy densities as inferred here, the presence of backsplash
galaxies returning to the apocenter of their orbit would help to
explain our results. Such galaxies comprise as much as half of all
galaxies beyond the virial radius in massive clusters
\citep{balo00}. The observed dependence on galaxy stellar mass arises
naturally under the assumption that low-mass galaxies are relatively
more gas-rich to begin with and have lower gravitational restoring
forces, making them more susceptible to gas removal.

\subsection{UV Luminosity Density in Groups}

As a final point, we consider the total UV luminosity density in the
local universe provided by groups similar to those in the {\em XI}
sample. This is relevant for understanding the contribution of these
environments to the local SFR density and hence the general importance
of the results presented. To first evaluate the space density of such
groups, we consider the catalog of \citet{eke04}. This is a
statistically more robust catalog of groups compiled from the entire
2dFGRS, of which the \citet{merc02} catalog, from which our groups
were drawn, is a subset. As the statistics of the group population in
the two catalogs are completely consistent \citep{rasm06b}, we can
reliably use the \citet{eke04} catalog for this exercise, benefiting
from its superior statistics. Within the full catalog (avoiding the
2dF survey edges and regions of poor completeness), we select groups
with $N_{\rm gal} \geq 5$, $\sigma \le 500$~km~s$^{-1}$, and
$z=0.05$--0.07, to mimic the {\em XI} selection criteria within a
sufficiently wide redshift interval to obtain useful statistics. For
the resulting total of 395 systems, we infer a space density of
$n_{\rm XI} = 3.37\pm 0.17\times 10^{-4}$~Mpc$^{-3}$ ($1\sigma$
statistical error).

Within $R=1$ Mpc of the 22 fully covered groups in the present sample,
we derive average specific rest-frame luminosities (prior to
dust-correction) of $3.7\times 10^{28}$ (NUV) and $2.3\times
10^{28}$~erg~s$^{-1}$~Hz$^{-1}$ (FUV) per group. With the space
density inferred above, this implies total UV luminosity densities of
$1.25\pm 0.06$ (NUV) and $0.77\pm 0.04 \times
10^{25}$~erg~s$^{-1}$~Hz$^{-1}$~Mpc$^{-3}$ for such groups.  These are
factors of 4.2 (4.6) below the total spatially averaged output in NUV
(FUV) at $z<0.1$ \citep{wyde05}, implying that {\em XI}--like groups
account for $24^{+6}_{-8}$\% ($22^{+5}_{-7}$\%) of the total NUV (FUV)
luminosity density in the local universe. In contrast to the
\citet{wyde05} results, these numbers do not incorporate corrections
for incompleteness below our $m=22.0$ limit and so should be
considered lower limits. Hence, despite the enhanced SF quenching in
groups relative to the field, at least $\sim 25$\% of all UV star
formation at low redshift still takes place within friends-of-friends
redshift-selected groups similar to the {\em XI} systems.

For the 22 groups combined, we further estimate a total FUV SFR of SF
galaxies within $R=1$~Mpc of 180~$M_\odot$~yr$^{-1}$, of which
54~$M_\odot$~yr$^{-1}$ (30\%) is unobscured. This corresponds to an
average $A_{\rm FUV} \approx 1.3$, in broad agreement with results for
both field \citep{wyde05} and cluster samples \citep{cort08b,hain11}.
This implies similar levels of mean UV obscuration for galaxies across
all types of global environment.

\section{Summary and Conclusions}\label{sec,summary}

Using {\em GALEX} imaging, we have presented dust-corrected star
formation rates and general UV properties of the galaxy population
within 23 redshift-selected groups at $z\simeq 0.06$. The data cover a
radius of 2.5~Mpc around each group, allowing us to establish UV
properties of group members from group cores into the general field,
down to an SFR completeness level of 0.06~$M_\odot$~yr$^{-1}$ before
dust corrections. Our conclusions may be summarized as follows.

The fractions of UV star-forming galaxies show a clear, systematic
decline toward the dense group cores. The fractions are suppressed by
35\% on average relative to the field, and the suppression is
significant out to at least $R \approx 1.5$~Mpc~$\approx
2R_{200}$. This implies that the group environment is modifying galaxy
properties out to similar overdensity radii as are much more massive
clusters \citep{balo98}. This remains true at fixed galaxy stellar
mass but is most pronounced for galaxies within the low-mass half of
our UV--detected sample ($M_\ast < 10^{9.8}$\,$M_\odot$). The
star-forming fractions within $R_{200}$ also decline with total group
stellar mass, from an average of $\sim 70$\% in the lowest-mass groups
to only $\sim 30$\% within the most massive systems ($\sigma \ga
400$~km~s$^{-1}$).

In a minority of the groups, the central galaxy is a late-type object
with NUV--R colors consistent with ongoing star formation. These
groups tend to be of lower total stellar mass and richness, and are
possibly dynamically younger than the rest, as also suggested by the
larger average projected distance between their central galaxy and the
group luminosity centroid.

The star formation rates of group galaxies depend primarily on galaxy
stellar mass, in a manner broadly agreeing with results for the global
low-redshift galaxy population, SFR\,$\propto M_\ast^{\sim 0.6}$.
There is no evidence for enhanced star formation at intermediate radii
or galaxy densities, as otherwise reported for galaxies in the
filaments connecting massive clusters. At stellar masses $M_\ast >
10^{8.5}$\,$M_\odot$, the global fraction of starburst galaxies within
the groups is $\approx 3$\%, consistent with, but at the low end of,
the range inferred globally for the nearby universe. The fraction
declines significantly with local galaxy density, however, from 5--7\%
in the lowest-density regions to $\la 1$\% in the dense group cores.

For the first time, we report a dependence on local galaxy environment
of not only star-forming galaxy fractions but also of specific SFRs
{\em within} blue star-forming galaxies at fixed $M_\ast$. The latter
trend is strongest for low-mass galaxies ($M_\ast \la
10^9$\,$M_\odot$), in which the mean specific SFR declines by a factor
$\approx 2.5$ from the field to the dense group cores. This effect
becomes insignificant at galaxy stellar masses $M_\ast \ga
10^{10}$\,$M_\odot$, plausibly explaining why it has not been detected
in other detailed studies of groups at higher redshift. On average,
specific star formation rates of star-forming galaxies in groups are
suppressed by $\approx 40$\% relative to the field, taking results for
the field either from our own analysis or from other {\em GALEX}
results for large, low-redshift galaxy samples.

At fixed galaxy mass and local density, the suppression of both
specific SFRs and star-forming galaxy fractions is stronger in more
massive groups, especially in the dense group cores. The decline in
both quantities with local density is in fact largely driven by the
high-mass groups with red central galaxies, whereas most galaxies in
lower-mass groups with blue centrals may have yet to be significantly
affected by their environment. Both local and global group environment
thus play a role in quenching star formation, suggesting a combination
of at least two underlying mechanisms: One whose efficiency depends
primarily on radius or local galaxy density, and one which scales with
total group mass and acts primarily in the group cores. Galaxy--galaxy
interactions and gas-dynamical processes, respectively, are the
obvious candidates.

The observed decline in specific SFRs toward group cores implies a
characteristic timescale for shutting down star formation which is
comparable to the crossing times, $\sim 2$~Gyr. This timescale is in
good agreement with other independent estimates. Based on the observed
starburst fraction in the groups, we demonstrate that gas consumption
associated with starbursts triggered by galaxy--galaxy interactions
proceed too quickly to be compatible with this timescale. Combined
with the absence of a bimodality in the distribution of specific SFRs,
this result confirms that a more slowly acting mechanism also
contributes significantly to quenching in groups. We argue that
starvation is the most likely candidate, acting in concert with
galaxy--galaxy interactions.

The generality of these results for the global low-redshift galaxy
population is attested by the finding that the total UV luminosity
density provided by typical groups with more than four members is at
least 25\% in the local universe. This is in spite of the accelerated
quenching of star formation relative to the field still taking place
within these systems at present. We also find that $\sim 70$\% of the
UV star formation in such group environments is obscured, in good
agreement with results obtained for field and cluster samples.

\acknowledgments 

We thank the referee for helpful and constructive suggestions which
improved the presentation of our results. We are grateful to Mark
Seibert and Anja von der Linden for useful discussions. This work has
made use of the NASA/IPAC (NED) extragalactic database. JR
acknowledges support provided by the Carlsberg Foundation, and by the
National Aeronautics and Space Administration through Chandra
Postdoctoral Fellowship Award Number PF7-80050. JSM acknowledges
partial support for this work from NASA grant NNX08AU586 and NSF grant
AST-0707417.\\

\end{document}